\documentclass[prl,aps,floats,floatfix,twocolumn,epsf,showpacs]{revtex4-1}
\usepackage[english]{babel}

\usepackage{amsmath,amssymb,amsfonts}

\usepackage{epsfig}
\usepackage[figtopcap,tight]{subfigure}
\usepackage[export]{adjustbox}

\usepackage{color}
\usepackage[dvipsnames]{xcolor}

\newcommand{\eq}[2]
{
  \begin{equation}
    #1
    \label{#2}
  \end{equation}
}

\newcommand{\eqsplit}[2]
{
  \begin{equation}
    \begin{split}
      #1
    \end{split}
    \label{#2}
  \end{equation}
}

\newcommand{\equ}[1]
{Eq.~(\ref{#1})}

\newcommand{\figu}[1]
{Fig.\ref{#1}}

\newcount\colveccount
\newcommand*\colvec[1]{
  \global\colveccount#1
  \begin{pmatrix}
    \colvecnext
  }
  \def\colvecnext#1{
    #1
    \global\advance\colveccount-1
    \ifnum\colveccount>0
    \\
    \expandafter\colvecnext
    \else
  \end{pmatrix}
  \fi
}

\newtoks\rowvectoks
\newcommand{\rowvec}[2]{%
  \rowvectoks={#2}\count255=#1\relax
  \advance\count255 by -1
  \rowvecnexta}
\newcommand{\rowvecnexta}{%
  \ifnum\count255>0
  \expandafter\rowvecnextb
  \else
  \begin{pmatrix}\the\rowvectoks\end{pmatrix}
  \fi}
\newcommand\rowvecnextb[1]{%
  \rowvectoks=\expandafter{\the\rowvectoks&#1}%
  \advance\count255 by -1
  \rowvecnexta
}

\def\bcen{\begin{center}}
\def\ecen{\end{center}}

\def\p{\pi}

\def\PP{{\cal P}}

\def\QQ{{\cal Q}}

\def\CC{{\cal C}}

\def\ZZZ{\mathbb{Z}}

\usepackage{bbold}
\def\11{\mathbb{1}}
\def\00{\mathbf{0}}

\def\=={\equiv}

\def\qed{\raise1pt\hbox{\vrule height5pt width5pt depth0pt}}

\def\cG0{{\cal G}_0}
\def\cG{{\cal G}}

\def\spinup{\uparrow}

\def\up{\uparrow}

\newcommand{\bfr}{{\bf r}}
\def\bra#1{\langle#1 |}
\def\ket#1{| #1\rangle}

\def\kx{{ k_x}}

\def\kxi{{k_x\!y}}

\def\yi{{ y}}
\def\xi{{ x}}
\def\ia{{\bf i}}
\def\ja{{\bf j}}

\def\opN{\hat{ N}}
\def\opSz{\hat{ S}_z}
\def\opTz{\hat{ T}_z}

 \def\Im{\mbox{Im}}

\def\=={\equiv}

\def\Im{{\rm Im}}
\def\Re{{\rm Re}}

\def\ep0{\epsilon_{p}}
\def\ed0{\epsilon_{d}}

\begin{document}
\title{Edge states reconstruction from strong correlations in quantum spin Hall insulators}

\author{A.~Amaricci}
\affiliation{Scuola Internazionale Superiore di Studi Avanzati (SISSA)
and Democritos National Simulation Center,
Consiglio Nazionale delle Ricerche,
Istituto Officina dei Materiali (IOM),
Via Bonomea 265, 34136 Trieste, Italy}
\author{L.~Privitera}
\affiliation{Scuola Internazionale Superiore di Studi Avanzati (SISSA)
and Democritos National Simulation Center,
Consiglio Nazionale delle Ricerche,
Istituto Officina dei Materiali (IOM),
Via Bonomea 265, 34136 Trieste, Italy}

\author{F.~Petocchi}
\affiliation{Scuola Internazionale Superiore di Studi Avanzati (SISSA)
and Democritos National Simulation Center,
Consiglio Nazionale delle Ricerche,
Istituto Officina dei Materiali (IOM),
Via Bonomea 265, 34136 Trieste, Italy}
\author{M.~Capone}
\affiliation{Scuola Internazionale Superiore di Studi Avanzati (SISSA)
and Democritos National Simulation Center,
Consiglio Nazionale delle Ricerche,
Istituto Officina dei Materiali (IOM),
Via Bonomea 265, 34136 Trieste, Italy}


\author{G.~Sangiovanni}
\affiliation{Institut f\"ur Theoretische Physik und
  Astrophysik, Universit\"at W\"urzburg, Am Hubland, D-97074 W\"urzburg, Germany}
\author{B.~Trauzettel}
\affiliation{Institut f\"ur Theoretische Physik und
  Astrophysik, Universit\"at W\"urzburg, Am Hubland, D-97074 W\"urzburg, Germany}

\date{\today}

\begin{abstract}
We study quantum spin Hall insulators with local Coulomb interactions in the presence of 
boundaries using dynamical mean field theory. We investigate the
different influence of the Coulomb interaction on the bulk
and the edge states. Interestingly, we discover an edge reconstruction 
driven by electronic correlations. The reason is that the 
helical edge states experience Mott localization for an interaction
strength smaller than the bulk one. We argue that the significance of this edge
reconstruction can be understood by topological properties of the
system characterized by a local Chern marker. 
\end{abstract}

\pacs{}

\maketitle

\paragraph{Introduction.--}
Topological insulators are symmetry-protected quantum materials with a
gapped bulk but gapless edge states. In two spatial dimensions, the
quantum spin Hall insulator (QSHI) is the prime example of a
topologically non-trivial phase of matter. Here, the underlying
symmetry that needs to be preserved is time-reversal symmetry (TRS)
\cite{Kane2005PRL,Kane2005PRLa,Bernevig2006S}. The QSHI phase is
usually detected by transport properties determined by its boundary
modes, which are coined
helical edge states \cite{Konig2007S,Roth09,Knez11} because their spin degree of freedom and their
direction of motion are strongly coupled to each other. 
This leads to a protection
from elastic backscattering off potential fluctuations
\cite{Zhang06,Moore06}. However, in experiments, edge state transport
in the QSHI is not perfect implying  some sort of backscattering
mechanism. The
interplay of Coulomb interaction and spin-mixing disorder has been
proposed as a possible origin of such inelastic backscattering
\cite{Strom2010,Budich12,Crepin12,Schmidt12,Glazman13,Geissler2014}. Thus,
it is crucial to better understand the influence of Coulomb
interaction on the physical properties of QSHIs. In fact, already for
bulk systems, i.e. in the absence of edge states, it has been shown
that topological insulators in the presence of strong Coulomb
interaction behave rather differently as compared to their weakly or
non-interacting counterparts
\cite{Lehur10,Hohenadler2011,Hohenadler2012,Tada2012PRB,Budich2013PRB,Hohenadler2013JOPCM,xHung2014PRB,xLu2013PRL,Amaricci2015PRL,Amaricci2016PRB,Roy2016PRB}. 


%
\begin{figure}
  \includegraphics[width=0.475\textwidth,center]{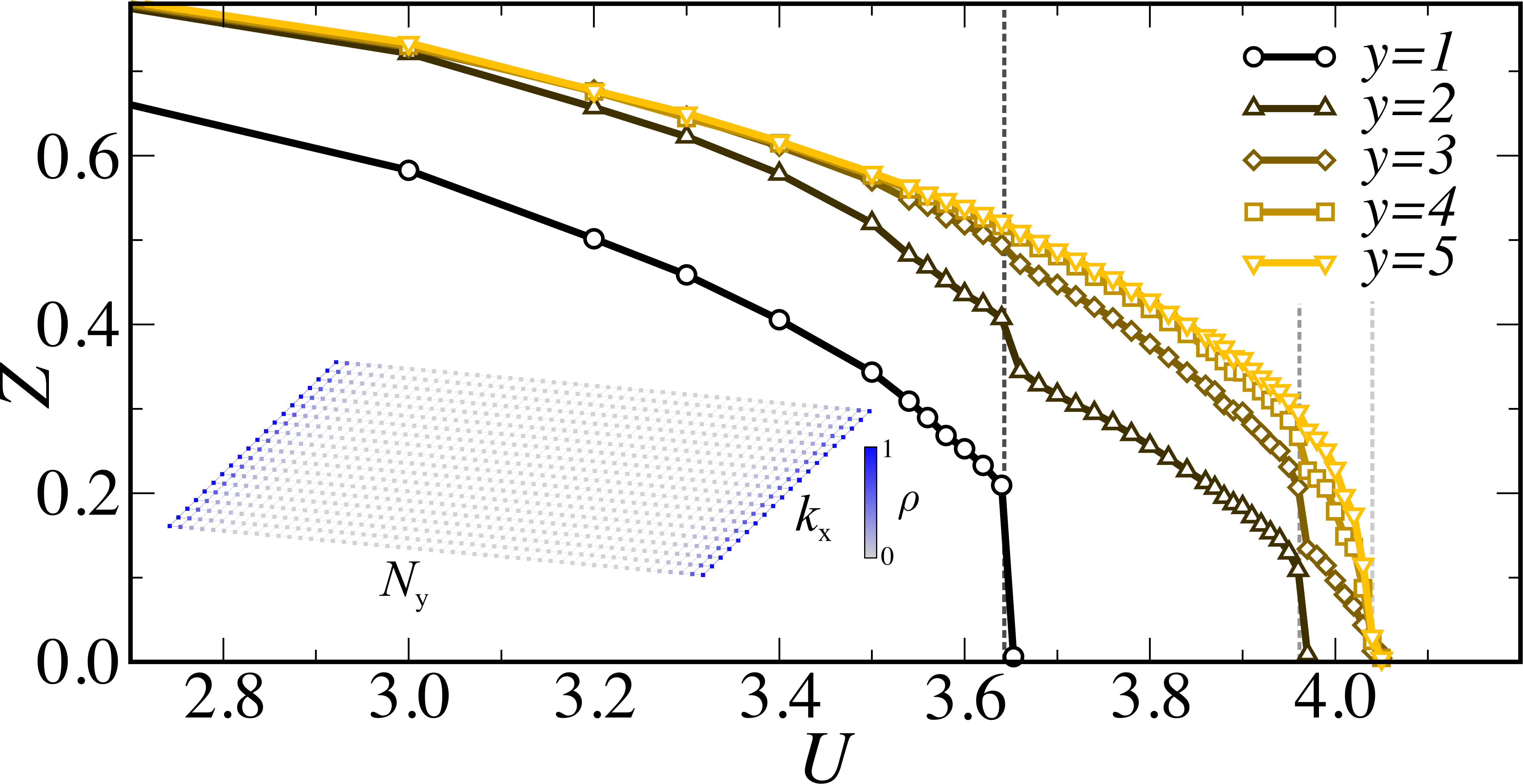}
  \caption{Quasiparticle weight $Z$ as a function of the interaction
    strength $U$ for the first few layers (under the parameter choice $M=1$ and $\lambda=0.3$ of the BHZ model, see Eq.~(\ref{model_hamiltonian}) below).
    The vertical dashed lines indicate the Mott
    localization of the first three layers, signaled by
    $Z_y=0$. Inset: schematic representation of the
    stripe system. Points of the lattice are colored according to the
    corresponding value of the spectral weight
    $\rho\!=\!A_{y}(\kx,\omega\!=\!0)$.}
  \label{fig1a}
\end{figure}
In this article we address the problem of how
the helical edge states in a strongly interacting QSHI phase are
affected by Coulomb interaction.
Using dynamical mean-field theory (DMFT) \cite{Georges1996RMP,Kotliar2006RMP}, we study the role of Coulomb
interaction in a paradigmatic model of the QSHI, called
Bernevig-Hughes-Zhang (BHZ) model \cite{Bernevig2006S}, in a stripe
geometry. We show that the influence of electronic correlation is more
pronounced at the boundaries than in the bulk. This perception is
nicely illustrated by the evolution of the quasiparticle weight $Z$,
which coincides with the inverse of the effective mass enhancement
within DMFT,  as  
a function of the interaction strength $U$ for the first few layers ($y=1$
being the outermost one), see Fig.~\ref{fig1a}. For non-topological
systems, it has been realized before that Mott localization can happen
more efficiently at the boundary of a given system than in the bulk
because the kinetic energy is effectively reduced at the edge
\cite{Borghi2009PRL}. However, for the QSHI this result
has a new implication. Since the Mott localization of the outer
helical edge state happens for a smaller interaction strength than
that of the bulk, the bulk itself remains in the QSHI phase. 
Hence, the Mott insulator at the boundary acts as a new
vacuum (still preserving TRS) resulting in a
new topological edge state that is moved inside the QSHI system. To
the best of our knowledge, this is the first example of a topological
edge reconstruction triggered by Coulomb interaction.

\paragraph{Model and Method.--}
We consider a two-orbital BHZ model supplemented by a local
interaction term~\cite{Budich2013PRB,Amaricci2015PRL}. 
The model is defined on a two-dimensional stripe system formed by a finite
number $N_y$ of one-dimensional layers stacked along the $y$ direction with open boundary conditions (OBC).
We assume translational invariance and periodic boundary conditions
(PBC) along the $x$ direction (see inset of \figu{fig1a}). 
%
We introduce the following $\Gamma$-matrices: 
$\Gamma_0 \!=\! \11 \otimes\11 $,
$\Gamma_x \!=\! \sigma_z \otimes \tau_x$,
$\Gamma_y \!=\!-\11 \otimes \tau_y $,
$\Gamma_5 \!=\! \11 \otimes \tau_z$,
$\Gamma_\sigma\!=\!\sigma_z\otimes\11$,
where $\sigma_{x,y,z}$ and $\tau_{x,y,z}$ are two sets of Pauli
matrices acting, respectively, on the spin and orbital sector, $\otimes$
is the usual tensor product, and $\11$ is the 2$\times$2 identity matrix.
Then, the model Hamiltonian reads
\eqsplit{
  H = &  \sum_{\kx,\yi,\yi'}
  \Psi^\dag_\kxi \mathbf{M}(\kx) \delta_{y y'} \Psi_{\kx\!y'} + \cr
  & \sum_{\kx\yi,\yi'}  \left(
  \Psi^\dag_\kxi \mathbf{T} \delta_{y\!+\!1 y'} \Psi_{\kx\!y'} \! +\!
  H.c.\right)
  + H_\mathrm{int} \; ,
}{model_hamiltonian}
where $\kx$ is the first component of the wave-vector, $\yi\!=\!1,\dots,N_y$ is the coordinate in the
$y$ direction (i.e. the layer index),
$\mathbf{M}(\kx)\!=(\!M\!-\!\epsilon\cos(\kx))\Gamma_5 \! +\!
\lambda\sin(\kx)\Gamma_x$, $  \mathbf{T}\!=\!-\frac{\epsilon}{2}\Gamma_5 +
i\frac{\lambda}{2}\Gamma_y$ and 
$\Psi_\kxi^\dag= ( c_{1\uparrow}^\dag ,\, c_{2\uparrow}^\dag ,\,
c_{1\downarrow}^\dag ,\, c_{2\downarrow}^\dag )_\kxi$ where the first index
refers to the orbital degree of freedom.
The first two terms in Eq.~(\ref{model_hamiltonian}) describe a system of
two bands of width $W\!=\!6\epsilon$, hybridized with an amplitude
$\lambda$ and separated in energy by a splitting of $2M$.
In the following, we consider a total density of two electrons per
site, i.e. a system at half-filling. We assume $\epsilon$ as
our energy unit and choose $\lambda\!=\! 0.3$ and $M\!=\!1$ (if not
stated differently).

The last term of the model Hamiltonian describes a local Coulomb interaction with both  {\it
  inter}- and {\it intra}-orbital repulsion and the Hund's coupling
$J$, taking into account the exchange effect which favors  high-spin configurations.
In terms of the local operators,
$\opN\! =\!\sum_{\ia\ja}\Psi^\dag_{\ia}\Gamma_0\delta_{\ia\ja}\Psi_{\ja}$,
$\opSz\!=\!\tfrac{1}{2}\sum_{\ia}\Psi^\dag_{\ia}\Gamma_\sigma\delta_{\ia\ja}\Psi_{\ja}$,
$\opTz\!=\!\tfrac{1}{2}\sum_{\ia}\Psi^\dag_{\ia}\Gamma_5\delta_{\ia\ja}\Psi_{\ja}$,
the interaction term reads:
\eq{
  H_\mathrm{int} = (U-J)\frac{\opN(\opN-1)}{2}
- J \left(  \frac{\opN^2}{4} + \opSz^2 - 2\opTz^2 \right)\, ,
}{model_interaction}
where $U$ is the strength of the electron-electron interaction and
 $\Psi_{\ia=\xi,\yi}=\sqrt{\tfrac{2\p}{V}}\sum_\kx e^{-i\kx\cdot\xi}\Psi_\kxi$.
\footnote{This Hamiltonian only contains the ``density-density'' part of the Hund's
exchange and neglects the so-called pair-hopping and spin-flip
terms. The robustness of the topological transitions in the BHZ
model against the pair-hopping and spin-flip terms has been
verified in Ref.~\onlinecite{Budich2013PRB}.}. In the remainder of the article,
we fix  $J\!=\!U/4$ but none of our results are specific to this choice.

We solve the interacting problem non-perturbatively using DMFT,
focusing on non-magnetic solutions. This choice implies that Coulomb
interaction will not lift the TRS protecting the QSHI phase. 
Anti-ferromagnetic ordering, which can be expected at low temperature
in the absence of frustration, would instead break TRS
leading to a different boundary scenario
beyond the scope of the present work.
 In order to capture the
different behavior between bulk and boundaries, we use an extension of
DMFT to treat inhomogeneous systems
\cite{yPotthoff1999PRB,yFreericks2006,yAmaricci2014PRA}.
In this framework, the interaction effect is contained in  
layer-dependent self-energy functions $\Sigma_y(\omega)$, bearing the
correct spin-orbital structure.
From this quantity we compute the layer-dependent quasiparticle weight
$Z_y\!=\!(1\!-\!\partial\Re\Sigma_y(\omega)/\partial\omega_{|_{\omega
    \rightarrow 0}}\!)^{-1}$, which coincides in DMFT with the inverse of the effective
mass enhancement and it is a direct measure of the localization effect induced by the
interactions. In non-interacting systems, $Z_y\!=\!1$, while
$0\!<\! Z_y\! <\! 1$ denotes systems with finite electronic correlations; $Z_y=0$ is the
hallmark of Mott localization.

\begin{figure}
  \includegraphics[width=0.5\textwidth]{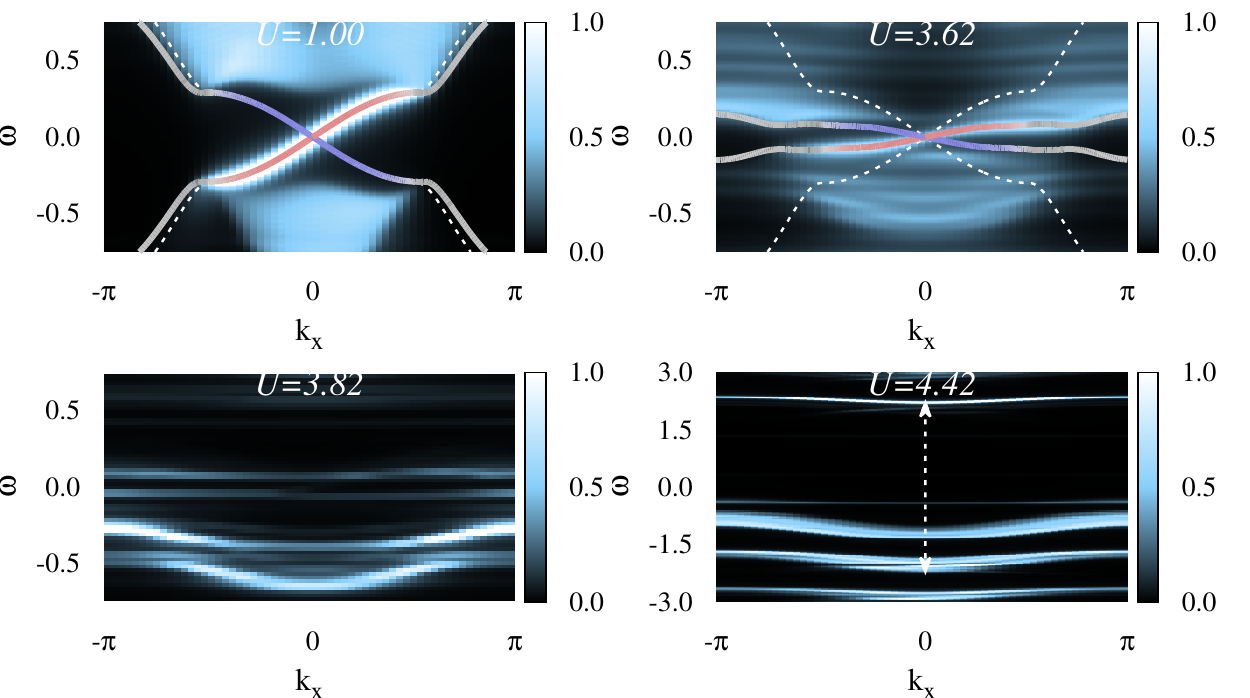}
  \caption{Evolution of the low-energy spectral function
    $A_{y}(\kx,\omega)$ (of the first orbital and spin up) for different interaction strength $U$.
    The dashed and solid (red and blue) lines indicate, respectively,
    the bare and the renormalized dispersion relation of the helical edge states.
    In the last panel the arrow indicates the width of the Mott gap.}
  \label{fig1b}
\end{figure}
\begin{figure*}
  \includegraphics[width=0.3295\textwidth]{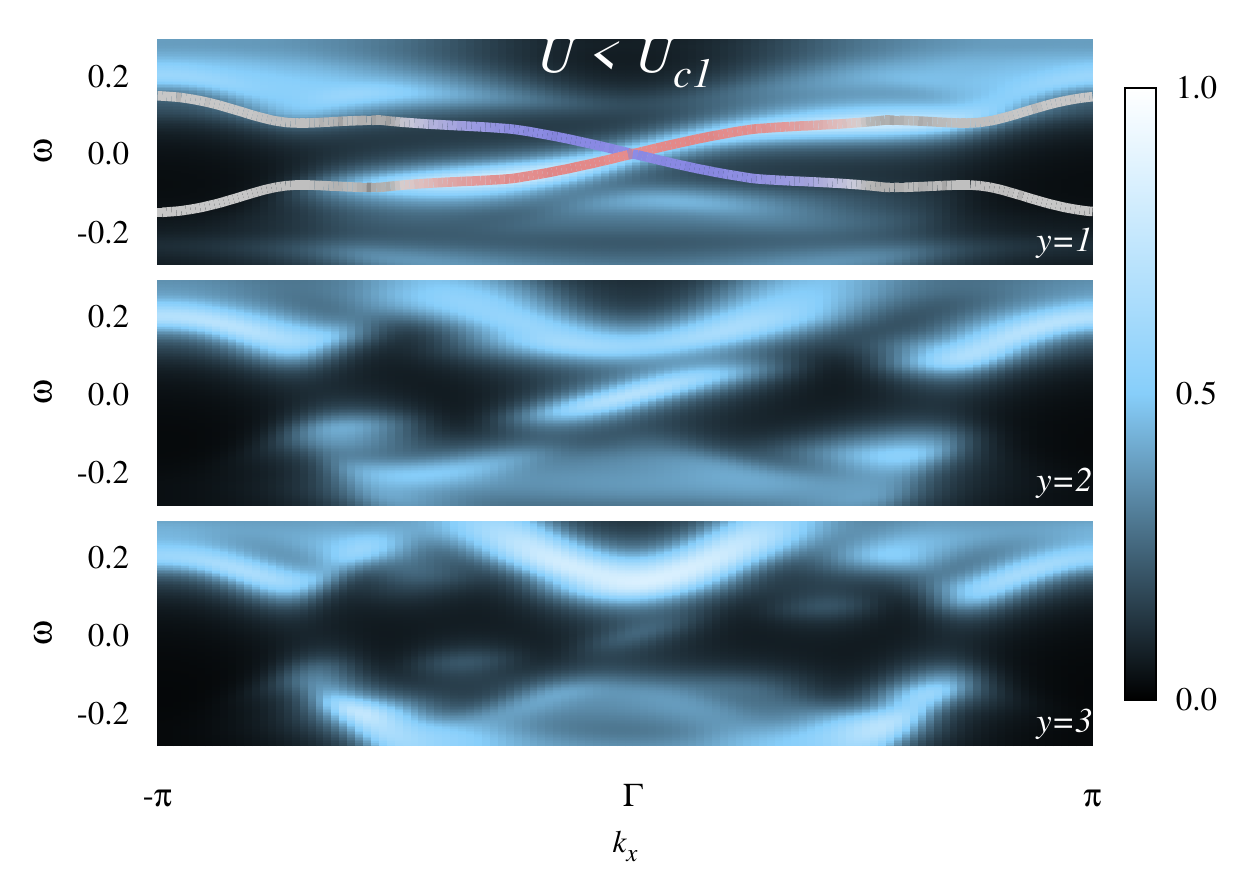}
  \includegraphics[width=0.3295\textwidth]{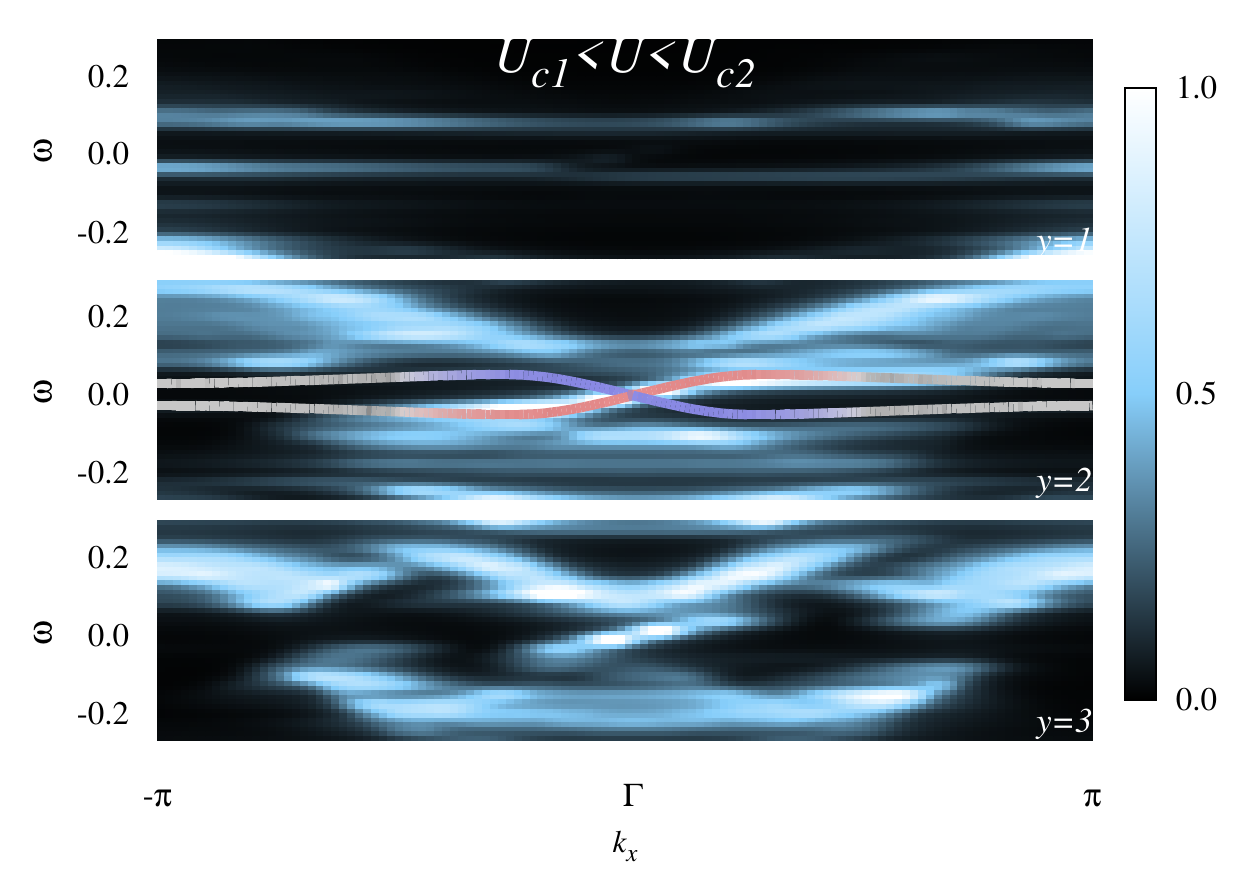}
  \includegraphics[width=0.3295\textwidth]{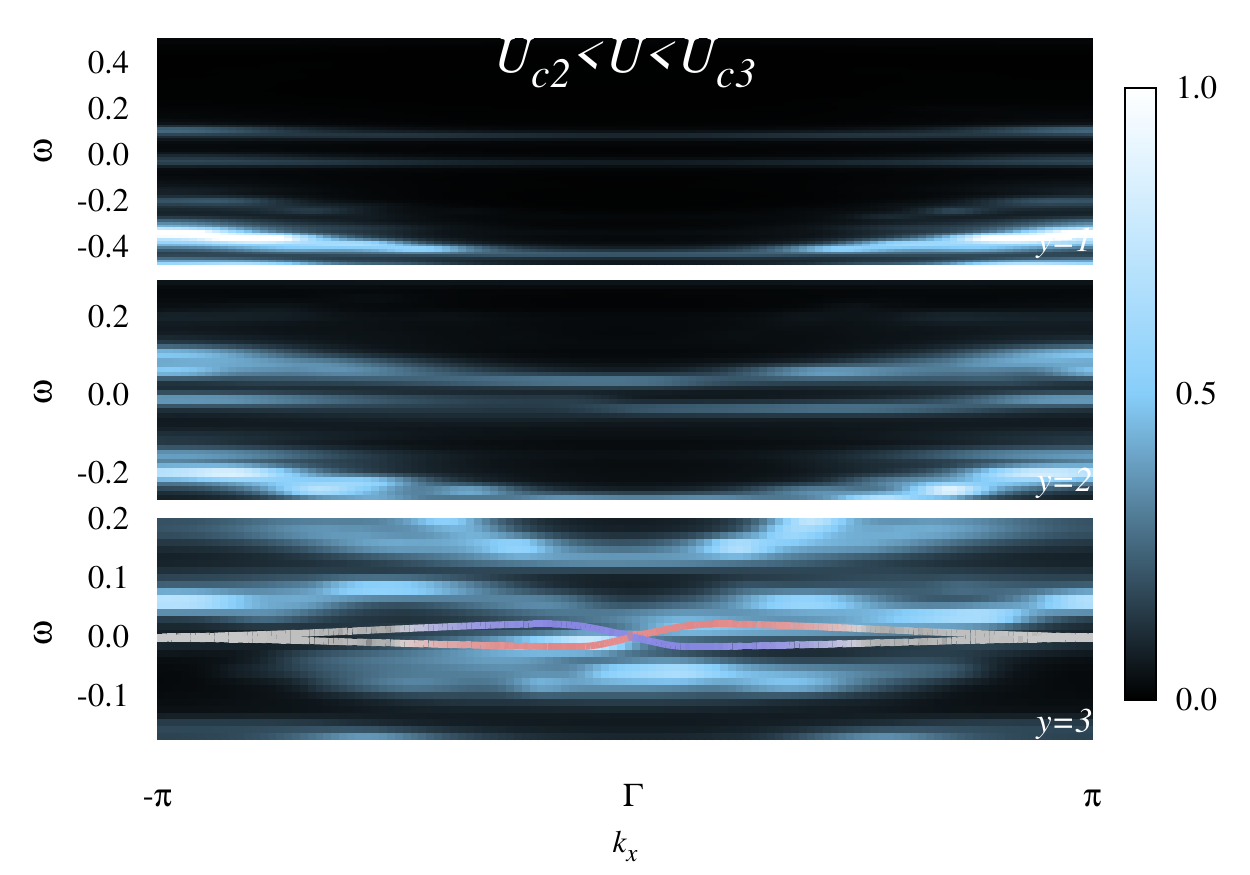}
  \caption{
    Evolution of the spectral functions $A_{y}(\kx,\omega)$ (of the
    first orbital and spin up) for the first three layers $y=1,2,3$
    (in the corresponding rows) as a function of the interaction
    strength    
    $U\!=\!3.62\!<\!U_{c1}$ (left), $U_{c1}\!<\!
    U\!=\!3.82\!<\!U_{c2}$ (center) and $U_{c2}\!<\!
    U\!=\!3.98\!<\!U_{c3}$ (right). For clarity of presentation, we
    picked just one orbital and one spin degree of freedom. The
    limited spatial extent of the edge states of about three layers
    can be nicely seen. The solid lines (red and blue) indicate the
    renormalized helical edge states dispersion (depicted for all
    spin-orbital channels to emphasize their helical character).    
  }
  \label{fig2}
\end{figure*}
\paragraph{Results.--}
In the non-interacting regime, $U\!=\!J\!=\!0$, the BHZ model (in its lattice-regularized version) describes a topological
quantum phase transition (TQPT) for $M\!=\!2$, separating a trivial
Band Insulator (characterized by a topological invariant $\nu\! =\!0$) for $M>2$ from a QSHI (with topological invariant $\nu\! =\!1$) for
$M<2$ \cite{Budich2013PRB}. In the QSHI phase the model hosts helical
edge states localized at the two boundaries of the stripe, as shown in
the inset of \figu{fig1a}, where we report the zero-frequency
local spectral weight for the non-interacting system, which is finite for
the edge states and vanishes in the bulk. 
In the bulk, the combined effect of the interaction terms $U$ and $J$ is to favor
the equal population of the two bands, effectively reducing the
energy splitting $M$~\cite{Budich2013PRB,Amaricci2015PRL}. 
This favors the QSHI over the trivial insulator, but it
also changes the character of the transition in the strongly
interacting regime from continuous to first-order \cite{Amaricci2015PRL,Amaricci2016PRB,Roy2016PRB}.
Finally, a large enough interaction strength drives the transition to a 
topologically trivial high-spin Mott
state~\cite{Amaricci2015PRL,Amaricci2016PRB}

In this work, we consider the stripe geometry introduced above. 
In this geometry, the correlation effects acquire a spatial modulation because of the existence of
boundaries, breaking the translational symmetry in the $y$ direction.
This can be understood noticing that the electrons at the
boundary layers can not hop in the outside direction. They have 
therefore a reduced kinetic energy with respect to electrons in the
bulk, which makes the interaction terms more effective at the edges.
This effect is shown in \figu{fig1a}, reporting the layer-resolved 
quasiparticle weight $Z_y$ 
as a function of the interaction strength $U$. 
%
In the weak-coupling regime ($U\!=\!1$), the effect of the Coulomb
interaction is to just renormalize the bands, without affecting the
qualitative nature of the QSHI solution. 
Upon increasing the interaction strength, we observe the progressive
reduction of the $Z_y$'s. 
The boundary value $Z_1$ is the smallest and $Z_y$ 
increases approaching the bulk layers. 
Most interestingly, $Z_1$ abruptly vanishes at  $U_{c1}\!\simeq\!3.64$, significantly smaller than
the bulk Mott transition point.  Further increasing the interaction strength, $Z_2$
vanishes abruptly at a second critical interaction  $U_{c2}>U_{c1}$,
eventually followed by the other layers. The successive critical points
$U_{cy}$, where $Z_y$ vanishes,  appear to rapidly accumulate to a
$U_c^{stripe}$ still smaller than the bulk critical interaction
strength (that we have previously
determined~\cite{Budich2013PRB,Amaricci2015PRL}). 
At the critical interaction strengths $U_{cy}$, all the inner layers show a little jump in their
quasiparticle weights. We can thus conclude that the value 
$U_c^{stripe}$, where all the quasiparticle
weights vanish, is the critical interaction strength for a full Mott localization of the whole QSHI stripe.

In order to get insight on the critical points $U_{cy}$ 
we study the evolution of the low-energy
part of the spectral function, which contains the single-particle
excitations of the correlated system. To avoid excessively busy figures, we plot only the contribution from
one orbital and one spin degree of freedom to $A_{y}(\kx,\omega) \!\equiv\!-\tfrac{1}{\pi}\Im
G_{y\; m\!=\! 1\; \spinup}(\kx,\omega\!+\!i0^+)$, because the other components contain exactly the same
physical information and can be reconstructed by symmetry.

In \figu{fig1b}, we report the evolution of the boundary spectral
function $A_{1}(\kx,\omega)$. The figure shows that the helical edge
states are renormalized by $Z_1$ for $U < U_{c1}$ and they
discontinuously disappear at $U_{c1}$, leaving behind a small
low-energy gap which turns into a large Mott-like gap of order $U$
only for values of the interaction strength $U>U_c^{stripe}$, see the
last panel in \figu{fig1b}.
Therefore, the jump of $Z_1$ at $U_{c1}$ can be described as a
selective localization transition in which the delocalized helical
edge states undergo some form of Mott localization, 
preceding the full Mott transition of the bulk.
Analogously, $U_{c2}$ marks a similar selective Mott localization for
the second layer, and so on.

In the intermediate regime, comprised between the gap opening of the
edge states and the Mott transition, the system opposes to the
strong interaction with a  {\it contraction} of the bulk and a {\it reconstruction}
of the helical edge states. This is the central result of our work.
To demonstrate this effect, we report in \figu{fig2} the
evolution of the spectral function $A_{y} (\kx,\omega)$ and the renormalized gapless edge states for the first few layers across
the multiple transition points indicated in \figu{fig1a}.
For small values of the interaction strength ($U\!=\!3.62\! <\!U_{c1}$, left column), the
system is characterized by the presence of helical edge
states localized at the outermost layer and a gapped bulk.
Increasing the interaction strength above the first critical point
($U_{c1}\!<\! U\! = \!3.82\! <\! U_{c2}$, middle column), this opens a gap
in that first layer.
This process is accompanied by the formation of a new pair of
helical edge states in the neighboring internal layer as to preserve the topologically
non-trivial character of the system against Coulomb interaction, according to the bulk-boundary correspondence.
Further increasing the interaction strength ($U_{c2}\!< \!U\! =\!
3.98\! <\! U_{c3}$, right column) causes a gap opening also in the
second layer with the collapse of the gapless helical states which
again are reconstructed in the adjacent layer. The bulk contraction
and edge state reconstruction process end when the interaction
strength becomes too large to prevent the reformation of renormalized
helical edge states. At this point, a large Mott gap penetrates the
system, transforming it into a trivial Mott insulator. 
We next substantiate our explanation of the edge reconstruction by a topology analysis.

\paragraph{Topological Invariant.--}
The topological nature of the QSHI is described  by the global
invariant $\nu\in\ZZZ_2$ \cite{Kane2005PRL,Kane2011PW}.
In presence of TRS and conserved spin $\nu\!=\!\
(\CC_\uparrow\! -\! \CC_\downarrow)/2\; \mathrm{mod}\, 2$, where
$\CC_{\sigma}$ is the Chern number for a given spin channel $\sigma$.
The value of $\CC_\sigma$ is usually expressed as a suitable
integral over the Brillouin zone \cite{Wang2012PRX,Hohenadler2013JOPCM}. However,
the topological character can also be obtained in real
space (Wannier) representation using a different, yet
equivalent, formulation originally introduced by Bianco and Resta \cite{Bianco2011PRB,
  Bianco2014chern}, which is particularly
useful in presence of OBC.
The key idea is to rewrite the Chern number in real space as \cite{PriviteraPRB16}
\eq{
\CC_{\sigma} = \lim_{V\to\infty} \frac{1}{V} \int_{V} \!d\bfr \;  2\pi i
\bra{\bfr}
\left(
\hat{x}^\sigma_{\PP} \hat{y}^\sigma_{\QQ} -
\hat{y}^\sigma_{\PP} \hat{x}^\sigma_{\QQ}
\right)
\ket{\bfr}
}{Clocal}
where $\hat{x}^\sigma_{\PP}=\PP^\sigma \hat{x}
\QQ^\sigma$ and $\hat{y}^\sigma_{\QQ}=\QQ^\sigma \hat{y} \PP^\sigma$
are the projected position operators,  and $\PP^\sigma$ and
$\QQ^\sigma$ are, respectively, the projectors onto the occupied and
empty bands of spin $\sigma$ of the topological
Hamiltonian \cite{Wang2012PRX}. The integrand of
\equ{Clocal} defines the local Chern marker (LCM) $C_\sigma(x,y)$ \cite{Bianco2011PRB,
  Bianco2014chern}. For a finite system the right hand side of
Eq.~(\ref{Clocal}) exactly vanishes.
However, the {\it nearsighted} nature
\cite{Kohn1996PRL,Resta2011TEPJB} of the electronic wave functions
guarantees that the LCM coincides with the Chern number in the bulk
of a sufficiently large system.
Deviations from this value can appear at the boundary of the 
non-trivial sample to insure the vanishing of the integral
(\ref{Clocal})~\footnote{The integral
  (\ref{Clocal}) can vanish either trivially, because the integrand is
  identically zero as for the Mott insulator, or in a non-trivial way
  as for the QSHI.}.
Thus, all the information about the topological nature of our
stripe system is provided by the LCM $C_\sigma(x,y)$
\cite{Bianco2011PRB,TranPRB15,PriviteraPRB16}. 

\begin{figure}
  \includegraphics[width=0.5\textwidth]{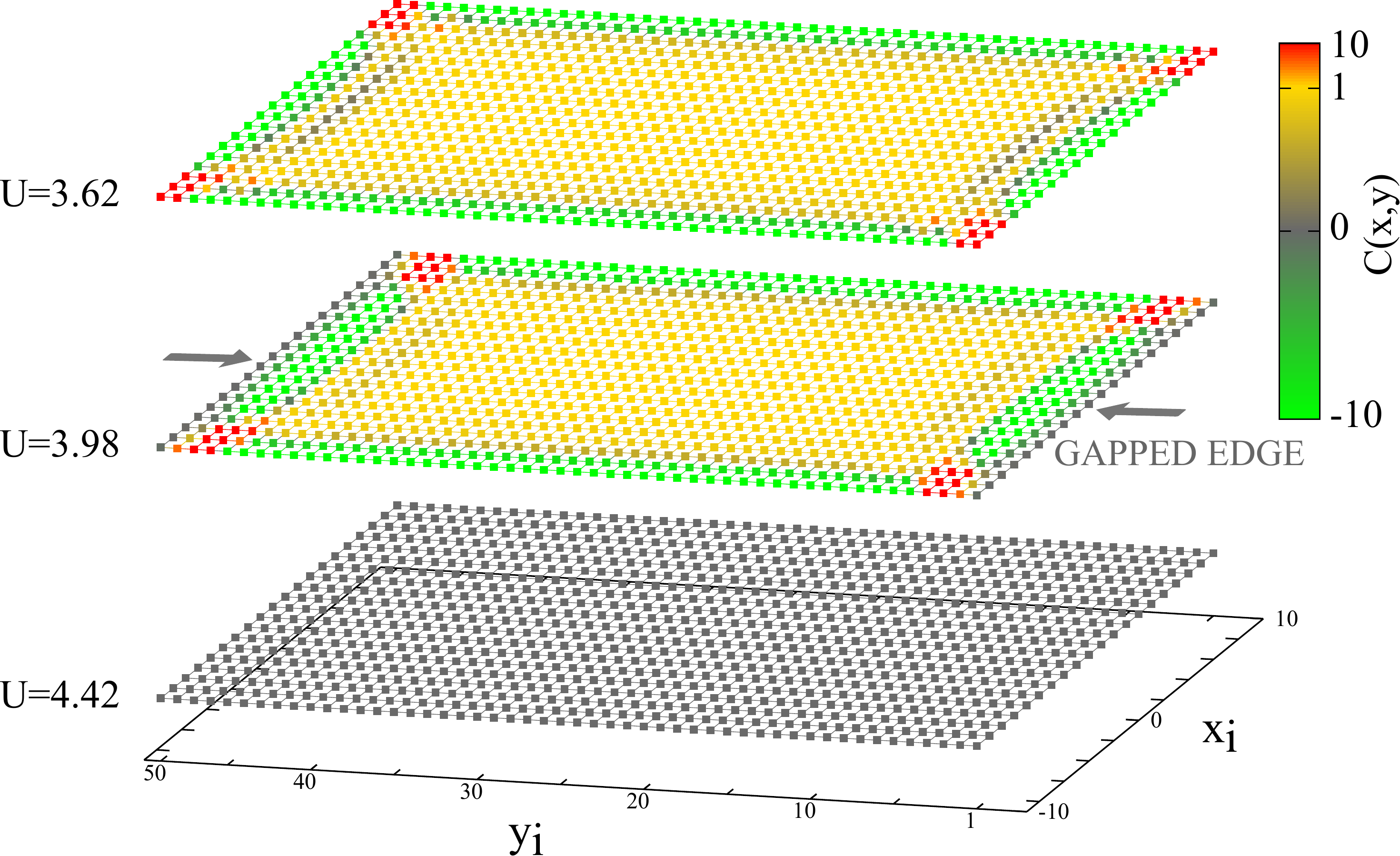}
  \caption{
    Spatial distribution of the Chern marker $C_{\sigma\!=\!\up}(x,y)$ for different values of the
    interaction strength $U$. Data is shown for a finite lattice with dimensions
    $N_x\!\times\!N_y\!=\!20\!\times\!50$ with PBC along $x$ and OBC
    along $y$. Each point of the system is colored according to the
    value of $C_\up$. 
    Top layer: The marker has constant value $C_\up\!=\!1$ in the bulk. 
    Oscillations at the border are a feature of the Chern marker in
    the non-trivial phase, but do not appear in the trivial one.
    Middle layer: The marker is $C_\up\!=\!1$ in the bulk, but gapped
    edge states at $y\!=\!1,50$ have $C_\up\!=\!0$. 
    Bottom layer: The marker is identically zero in the trivial Mott state.
  }
  \label{fig3}
\end{figure}

A crucial prerequisite of the observed edge reconstruction (driven by Coulomb interaction) is the preservation of the non-trivial topological
character in the interior part of the stripe. The reconstructed helical edge states then separate the QSHI in the spatially shrunken
bulk from the Mott gapped external layers. In order to illustrate this point, we evaluate the LCM $C_\sigma(x,y)$, with
$x$ the coordinate conjugated to $\kx$ in a finite lattice of
dimension $N_x\!\times\!N_y$.
The bulk averages of $C_\sigma(x,y)$ give the $\ZZZ_2$ 
topological invariant characterizing the QSHI state. 
The evolution of the LCM distribution across the Mott
transition is reported in \figu{fig3}.
In the non-trivial phase, the $C_\sigma(x,y)$ undergo positive
and negative oscillations at the boundary which cancel out the bulk
contribution to the integral (\ref{Clocal}), see top panel in \figu{fig3}.
However, such oscillations are not present in the trivial phase (see bottom panel).
This enables us to discriminate between trivial and non-trivial parts of
the same system.
Indeed in the intermediate regime preceding the full
Mott localization, the LCM of the external layers becomes
zero whereas the non-trivial part with $\nu\!=\!1$ is compressed towards the
bulk. 
%

\paragraph{Conclusions. -- }
We have studied the properties of the edge states in a BHZ model in the presence of
strong (on-site) Coulomb interactions.
Our analysis shows that the correlation-driven transition from a QSHI
to a Mott insulator occurs through a series of selective localization
transitions as the interaction grows, before the bulk eventually
turns into a Mott insulating state. At each localization transition the outermost helical
edge states are gapped by the interaction.
The disappearance of the helical modes at the outer layer is accompanied by a reconstruction
of new (renormalized) edge states at the immediately inner layers. This phenomenon demonstrates the
survival of a topologically non-trivial QSHI state in the bulk -- a direct consequence of the bulk-boundary correspondence.
Our results show that novel physics can be seen in theoretical approaches able to treat simultaneously bulk and boundaries of
correlated topological materials.

\paragraph{Acknowledgments. -- } We thank J.C. Budich for interesting discussions and a critical reading of the manuscript.
A.A., F.P., and  M.C. acknowledge financial support from MIUR through the PRIN 2015 program (Prot. 2015C5SEJJ001), the Seventh Framework
Programme FP7, under Grant No. 280555 ``GO FAST'', and the H2020 Framework Programme, under ERC Advanced Grant No. 692670 ``FIRSTORM''.
G.S. and B.T. acknowledge financial support by the DFG (SPP 1666 on ``Topological Insulators'' and SFB 1170 ``ToCoTronics'').

\bibliography{references}
\end{document}